# CRYPTANALYSIS OF A MORE EFFICIENT AND SECURE DYNAMIC ID-BASED REMOTE USER AUTHENTICATION SCHEME

Mohammed Aijaz Ahmed[1], D. Rajya Lakshmi[2]  and Sayed Abdul Sattar[3]

[1]Department of Computer Science and Engineering, GITAM University, Vishakapatnam
mohd_aijaz@yahoo.com
[2]Department of Information Technology, GITAM University, Vishakapatnam
rdavuluri@yahoo.com
[3]Department of Computer Science and Engineering, J.N.T. University, Hyderabad
syed49in@yahoo.com

## ABSTRACT

*In 2004, Das, Saxena and Gulati proposed a dynamic ID-based remote user authentication scheme which has many advantage such as no verifier table, user freedom to choose and change password and so on. However the subsequent papers have shown that this scheme is completely insecure and vulnerable to many attacks. Since then many schemes with improvements to Das et al's scheme has been proposed but each has its pros and cons. Recently Yan-yan Wang et al. have proposed a scheme to overcome security weaknesses of Das et al.'s scheme. However this scheme too is vulnerable to various security attacks such as password guessing attack, masquerading attack, denial of service attack.*

## KEYWORDS

*Password, Authentication, Smartcard, Remote User, Masquerade Attack*

## 1. INTRODUCTION

In order to prevent the invasion of privacy and solve security problems, a remote user would need to provide proof to a system that he/she is a legitimate user before he/she logs onto the remote system. There are many methods proposed to verify the legitimacy of a remote user such as password, fingerprint, typing sequence, and so forth. Among them, password based remote user authentication is extensively used and easily implemented to authenticate a legitimate user. In 1981, Lamport [3] proposed a password based authentication scheme that could authenticate remote users over a insecure channel. Since than many schemes [6, 7, 8, 9] have been proposed to improve security, efficiency, and cost.

In 2004, Das, Saxena and Gulati proposed a dynamic ID-based remote user authentication scheme [2] which has many advantage such as no verifier table, user freedom to choose and change password and so on. However the subsequent papers [4,5] have shown that this scheme is completely insecure and vulnerable to many attacks. Since then many schemes[1,4] with improvements to Das et al's scheme has been proposed but each has its pros and cons. Recently Yan-yan Wang et al. [1] have proposed a scheme to overcome security weaknesses of Das et al.'s scheme.

However in this paper we state that this scheme too is vulnerable to few security attacks such as password guessing attack, masquerading attack, denial of service attack.

The rest of the paper is organized as follows. In section 2 we present review of Yan-yan Wang et al.'s scheme. The section 3 describes cryptanalysis of Yan-yan Wang et al.'s scheme. And finally some concluding comments are included in the last section.

## 2. Review Of Yan-Yan Wang et al.'s Scheme

The scheme consists of four phases, the registration phase, the login phase, the verification phase and password change phase. The notations used in the scheme are as follows:

| | |
|---|---|
| U | The user |
| PW | The password of U |
| ID | The identity of U |
| S | The remote server |
| h(.) | A one-way hash function |
| ⊕ | Bitwise XOR operation |
| → | A common channel |
| ==> | A secure channel |
| A → B: M | A sends M to B through common channel |
| A ==> B: M | A sends M to B through secure channel |

### 2.1. Registration Phase

The user Ui sends the registration request to the remote server S:

1) Ui submits IDi to S

2) S computes:

$$Ni = h(PWi) \oplus h(x) \oplus IDi$$

Where x is secret of remote server, PWi is the password of Ui chosen by S.

3) S personalizes the smartcard with the parameters [h(.), Ni, y ], where y is the remote server's secret  number stored in each registered user's smartcard.

4) S ==> Ui: PWi and smartcard.

### 2.2. Login Phase

When a user wants to login the remote server, he/she inserts the smart card to the terminal and keys the identity IDi and the password PWi , then the smartcard performs the following steps:

1) Computes dynamic ID:

$$CIDi = h(PWi) \oplus h(Ni \Theta y \Theta T) \oplus IDi$$

Where T is the current date and time.

2) Ui → S: IDi, CIDi,Ni,T

## 2.3. Verification Phase

When the remote server S receives the request (IDi,CIDi,Ni,T) at time T' , S verifies as:

1) checks the validity of time interval, if $T' - T \leq \Delta T$ holds, S accepts the login request of Ui, otherwise the login request will be rejected, where $\Delta T$ is valid time interval.

2) S computes:

$$h'(PWi) = CIDi \oplus h(Ni \ominus y \ominus T) \oplus IDi$$

3) and computes

$$IDi' = Ni \oplus h'(PWi) \oplus h(x)$$

and verifies whether it is equal to IDi in the login request of Ui. If it does not hold S rejects the login request of Ui, otherwise accepts it. Then S computes a' using the result of step 2.

$$a' = h (h'(PWi) \oplus y \oplus T')$$

4) S sends (a',T) to Ui.

Upon receiving the reply message (a',T) at time T*, Ui verifies as:

5) Ui checks whether $T* - T' \leq \Delta T$, if it does then Ui computes $a = h(h(PWi) \oplus y \oplus T' )$ , and compares it with the received a' , if it holds, Ui confirms that S is valid.

## 2.4. Password Change Phase

When the user wants to change the password, he/she inserts the smartcard into the terminal device, keys the password PWi and request to change the password to new one PWnew, then the smartcard computes:

$$Ni* = Ni \oplus h(PWi) \oplus h(PWnew),$$

and replaces the Ni with the new Ni*, password gets changed.

## 3. CRYPTANALYSIS OF YAN-YAN WANG ET AL.'S SCHEME

In this section we will show that Yan-yan Wang et al.'s scheme is vulnerable to masquerade attack, password guessing attack, denial of service attack. Although tamper resistant smartcard widely assumed in most of the authentication scheme, but such an assumption is difficult in practice. Many researchers have shown that the secret stored in a smartcard can be breached by analyzing the leaked information or by monitoring the power consumption [10,11]. An attacker can extract secret y stored in the Ui's smartcard either by stealing the smartcard or by registering to the server (as each registered user has same value of y stored in their smartcard).

## 3.1. Password guessing attack

Assuming that the attacker has extracted the secret y from Ui's smartcard and also he/she has the intercepted parameters, CIDi, Ni, T and IDi. Then the attacker can proceed as follows:

$$h(PWi) = CIDi \oplus h(Ni \oplus y \oplus T) \oplus IDi$$

Now attacker can guess different passwords until the hash value of the guessed password matches with h(PWi) computed by the attacker.

## 3.2. User Masquerade Attack

In the second step of registration phase S computes:

Ni = h(PWi) ⊕ h(x) ⊕ IDi

The attacker can now extract h(x) from Ni by using h(PWi) computed in '*A*' :

h(x) = h(PWi) ⊕ Ni ⊕ IDi

now attacker can calculate new Ni* with his/her choosen password PW* as follows:

Ni* =  h(PWi*) ⊕ h(x) ⊕ IDi

Attacker can now create and send a forged login request to the remote server S, without knowing the original password:

CIDi* = h(PWi*) ⊕ h(Ni* ⊕ y ⊕ T*) ⊕ IDi

where T* is fresh time stamp.

Attacker sends to the server S, {CIDi*,Ni*,T*, IDi}. Upon receiving login request Server S successfully verifies validity of timestamp T *and identity IDi, hence accepting the request.

## 3.3. Server Masquerade Attack

The attacker can masquerade server by using the h(PWi) computed in ' *A* ' and :

a* = h (h(PWi) ⊕ y ⊕ T" )

Attacker then sends (a*,T" ) to Ui, which the user successfully verifies.

## 3.4. Denial of Service Attack

The password change phase of Yan-yan Wang et al.'s scheme is same as that of Das et al.'s scheme and it has a serious weakness. The password change phase does not verify whether the input old password matches with the original password. An attacker can use Ui's smartcard in his absence and can invoke password change phase by inputting an arbitrary password PW' in place of original password PWi along with a new password PWnew. Then the smartcard updates Ni without verifying the old password as follows:

Ni* = Ni ⊕ h(PW' ) ⊕ h(PWnew)

That will result in some arbitrary value Ni*.Now the original user Ui can not log onto the remote server even by using his correct password as the Ni has been changed to some arbitrary value.

## 3. CONCLUSION

In this paper, we briefly reviewed Yan-yan Wang et al.'s scheme and shown that this improved scheme too is vulnerable to various security attacks such as password guessing attack, user masquerade attack, server masquerade attack, denial of service attack. In addition to this the password change phase updates smartcard parameter even if wrong password is given as input.

**Authors**

[1]Md. Aijaz Ahmed received his B.E. Degree in Computer Science & Engineering from, M.B.E.S' College of Engineering, Ambejogai, Maharashtra, India in 2003; He has obtained M.E. in Computer Science & Engineering from, M.G.M's College of Engineering, S.R.T.M. University, Nanded, Maharashtra, India. He is currently pursuing Ph.D. in Computer Science & Engineering from GITAM University Vishakapatnam, Andhra Pradesh, India. His area of interest includes Network Security and Cryptography, Discrete Mathematics, Automata Theory.

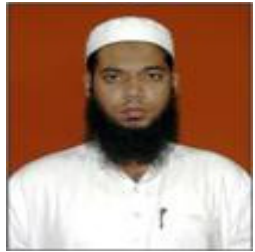

[2]Dr. D. Rajya Lakshmi is working presently as professor in Department of Information Technology at GITAM University, Visakhapatnam, AP, INDIA. Professor Rajya Lakshmi was awarded Ph.d in CSE from JNTU, Hyderabad. She has 16 years of teaching experience. Her research areas includes Image processing, Data mining, Network security.

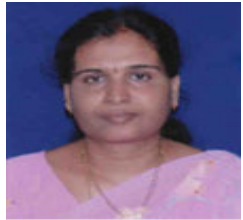

[3]Dr. Syed Abdul Sattar received B.E. (Electronics) from Marathwada University, Aurangabad, Maharashtra, India , in 1990. He received M.Tech. in Digital system and Computer Science from J.N.T. University, Hyderabad, Andhra Pradesh, India, in 2002. He received Ph.D. in Electronics & Communication Engg. from J.N.T. University, Hyderabad, in 2007. His area of interest include Computer Communications, Network Security, Image Processing.